\documentclass{jfm}

\usepackage{amsmath}
\usepackage{amssymb}
\usepackage{graphicx}
\usepackage{natbib}

\ifCUPmtlplainloaded \else
  \checkfont{eurm10}
  \iffontfound
    \IfFileExists{upmath.sty}
      {\typeout{^^JFound AMS Euler Roman fonts on the system,
                   using the 'upmath' package.^^J}%
       \usepackage{upmath}}
      {\typeout{^^JFound AMS Euler Roman fonts on the system, but you
                   dont seem to have the}%
       \typeout{'upmath' package installed. JFM.cls can take advantage
                 of these fonts,^^Jif you use 'upmath' package.^^J}%
       \providecommand\upi{\pi}%
      }
  \else
    \providecommand\upi{\pi}%
  \fi
\fi

\ifCUPmtlplainloaded \else
  \checkfont{msam10}
  \iffontfound
    \IfFileExists{amssymb.sty}
      {\typeout{^^JFound AMS Symbol fonts on the system, using the
                'amssymb' package.^^J}%
       \usepackage{amssymb}%
         
         \let\geq=\geqslant
      }{}
  \fi
\fi

\ifCUPmtlplainloaded \else
  \IfFileExists{amsbsy.sty}
    {\typeout{^^JFound the 'amsbsy' package on the system, using it.^^J}%
     \usepackage{amsbsy}}
    {\providecommand\boldsymbol[1]{\mbox{\boldmath $##1$}}}
\fi

\newcommand\de{\mathrm{d}}

\providecommand\bcdot{\boldsymbol{\cdot}}

\newcommand\img{\mathrm{i}}
\newcommand\e{\mathrm{e}}

\newcommand\Real{\mbox{Re}} 
\newcommand\Imag{\mbox{Im}} 
\newcommand\phir{\phi^{(R,D)}}               

\newsavebox{\astrutbox}
\sbox{\astrutbox}{\rule[-5pt]{0pt}{20pt}}

\newcommand\phijr{\ensuremath{\Delta\varphi^{(R,D)}_n}}             
\newcommand\Ha{\ensuremath{H_0^{(1)}}}							                       
\newcommand\Haa{\ensuremath{H_1^{(1)}}}						                       

\newcommand\p{\ensuremath{\partial}}

\newcommand\shalf{\ensuremath{{\scriptstyle\frac{1}{2}}}}

\newcommand\hint{\ensuremath{\int%
  \mskip \ifCUPmtlplainloaded -13mu\else -16mu\fi \times}}
\newcommand\hintw{\ensuremath{\int_{-w/2}^{w/2}%
  \mskip \ifCUPmtlplainloaded -44mu\else -47.5mu\fi \times\quad\quad}}
\newcommand\hintu{\ensuremath{\int_{-1}^{1}%
  \mskip \ifCUPmtlplainloaded -26mu\else -29mu\fi \times\quad}}

\title[Resonant behaviour of an OWEC in a channel]{Resonant behaviour of an oscillating wave energy converter in a channel}

\author[E. Renzi and F. Dias]%
{E\ls M\ls I\ls L\ls I\ls A\ls N\ls O\ns R\ls E\ls N\ls Z\ls I$^1$%
  \thanks{Email address for correspondence: emiliano.renzi@ucd.ie} \and
F.\ns D\ls I\ls A\ls S$^{1,2}$}

\affiliation{$^1$UCD School of Mathematical Sciences, University College Dublin, Belfield, Dublin 4, Ireland \\
$^2$Centre de Math\'{e}matiques et de Leurs Applications (CMLA), Ecole Normale Sup\'{e}rieure de Cachan, 94235 Cachan, France}

\pubyear{}
\volume{}
\pagerange{}
\date{?; revised ?; accepted ?. - To be entered by editorial office}
\begin{document}

\maketitle

\begin{abstract}
A mathematical model is developed to study  the behaviour of an oscillating wave energy converter in a channel. During recent laboratory tests in a wave tank, peaks in the hydrodynamic actions on the converter occurred at certain frequencies of the incident waves. This resonant mechanism is known to be generated by the transverse sloshing modes of the channel. Here the influence of the channel sloshing modes on the performance of the device is further investigated. Within the framework of a linear inviscid potential-flow theory, application of the Green theorem yields a hypersingular integral equation for the velocity potential in the fluid domain. The solution is found in terms of a fast-converging series of Chebyshev polynomials of the second kind. The physical behaviour of the system is then analysed, showing sensitivity of the resonant sloshing modes to the geometry of the device, that concurs in increasing the maximum efficiency. Analytical results are validated with available numerical and experimental data.
\end{abstract}

\begin{keywords}
Authors should not enter keywords on the manuscript, as these must be chosen by the author during the online submission process and will then be added during the typesetting process (see http://journals.cambridge.org/data/\linebreak[3]relatedlink/jfm-\linebreak[3]keywords.pdf for the full list)
\end{keywords}

\section{Introduction}
Renewed interest in wave power extraction has arisen in the recent years, following the need for clean energy resources alternative to fossil fuels. As a consequence, a lot of work has been done recently to discover effective design layouts for wave energy converters \citep[see][for an extensive review of wave energy conversion techniques]{MC07, CR08}. The work of \citet{Fetal05} and \citet{Wetal07} on a number of different devices showed that an effective solution to harness energy from water waves is represented by oscillating wave energy converters. Their simplest form is that of a buoyant flap hinged at the bottom of the ocean and has been named Oscillating Wave Surge Converter (OWSC). Under the action of incoming incident waves, the flap is forced to a pitching motion about the hinge, that is eventually converted into energy with the use of a generator. The converter is naturally placed in shallow waters, where the pitching of the flap combines well with the amplified horizontal motion (surge) of the fluid particles (hence the name OWSC). During the wave tank tests of \citet{Wetal07}, \cite{AL08} and \cite{JO09}, peaks in the hydrodynamic actions on the converter occurred at certain frequencies of the incident waves. This resonant phenomenon is known to be due to the excitation of the transverse sloshing modes of the channel \citep{SKM87}. However, there is still uncertainty on how the resonant transverse waves interact with the device and affect its efficiency \citep{JO09}. Here we address this unresolved issue by developing a three-dimensional mathematical model of an oscillating wave energy converter, namely the OWSC, inside a channel. Although much analytical modelling has been done for a large number of wave energy converters (WECs) \citep[see][for a collection of theories on WECs]{FA02, ME05, MC07}, few analytical studies are available in the literature concerning OWSCs. 

Indeed, the hydrodynamics of flaps has been widely studied in the past due to their usage as wavemakers \citep[see][for a compendium of wavemaker theories]{DD91}, with \citet{MI70} being among the first to suggest the employment of a moving paddle not only as a wave generator, but also as a wave energy absorber. Later on, the hydrodynamic characteristics of flat plates have been studied by \citet{PM92, PM94, PM95} and more recently by \citet{EP96} for both submerged and surface-piercing plates. In particular, the work of \citet{PM92} is a  seminal example of the application of hypersingular integral equations to problems in water waves. Nevertheless, all these analytical studies deal with two-dimensional (2D) geometries. Hence they are inadequate to describe the complex three-dimensional (3D) behaviour of the wave field around the flap noticed in the experimental analysis of the OWSC. A non-linear 3D analytical model has been developed by \citet{ME94} and \citet{Setal97} for the flap gates of the barriers designed to protect Venice from flooding. \citet{ME94} and \citet{Setal97} showed that trapped modes exist near a barrier of oscillating flaps in a channel. Since the narrowly spaced flaps span the whole channel width, this system is closed and can resonate only subharmonically (i.e. non-linearly). The work of \citet{Setal97} aims at reducing the resonant effects that could lead to undesired large oscillations of the flaps, thus undermining the effectiveness of the barrier.

Going back to wave energy, a wise design of the OWSC should exploit the beneficial effect of resonance in enhancing the amplitude of oscillation of the flap, thus allowing to capture more wave power. For bodies of various geometries in a channel, \citet{ET85}, \citet{YS89}, \citet{LE92} and \citet{CH94} already showed that linear resonance mechanisms influence the radiation and diffraction properties significantly. This influence manifests itself by the presence of peaks in the hydrodynamic characteristics of the system, at periods corresponding to the resonance of symmetric transverse standing waves within the channel lateral walls. As already mentioned, similar dynamics has been shown to occur also during wave tank tests of the OWSC, rising the following questions: (i) how linear resonance of transverse waves is triggered for a typical OWSC configuration in a channel, (ii) how do the transverse sloshing modes interact with the OWSC and modify its performance, and (iii) to what extent the results obtained in an experimental wave tank can be applied to predict the behaviour of the device in the open ocean.

A mathematical model of the OWSC in a channel is developed in the following sections to address the above points. The fluid is assumed to be inviscid and incompressible, the flow irrotational, the perturbation time harmonic. In \S\ref{sec:anmod} the governing equations are scaled and linearised according to first-order potential flow theory. The hydrodynamic problem is then decomposed into a radiation and a diffraction problem, which are solved separately using the hypothesis of thin plate and eventually summed up to obtain the total potential. The solution is achieved with the integration of a hypersingular problem, resulting from the application of the Green theorem to the physical domain. In \S\ref{sec:disc} the hydrodynamic characteristics of the system, such as the excitation torque, added inertia torque and radiation damping are fully investigated and the efficiency of the OWSC is assessed. Analytical results agree well with available numerical and experimental data. It is shown that all the above quantities are characterised by the occurrence of spikes at the cut-off frequencies of the transverse sloshing modes. The latter are excited in the channel by the diffraction of the incident wave field at the edges of the flap and influence the performance of the OWSC. Finally, a parametric analysis reveals how the transverse modes can increase the efficiency of the device, depending on the geometry of the system. 
\section{Analytical model}\label{sec:anmod}
\subsection{Governing equations}\label{sec:goveq}
\begin{figure}
  \centerline{\includegraphics[width=12cm]{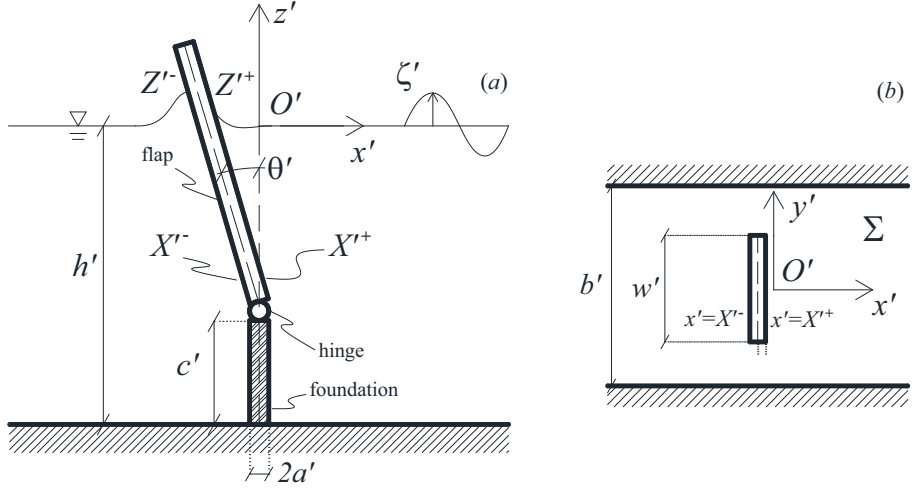}}
  \caption{Geometry of the converter and the channel: ($a$) section, ($b$) plan view.}
\label{fig:geom}
\end{figure}

Referring to figure \ref{fig:geom}($a,b$), consider an OWSC in water of depth $h'$, where the prime indicates a physical dimensional quantity. 
For the sake of generality, we shall represent the converter by a rectangular flap of width $w'$ and thickness $2 a'$, hinged along a straight axis upon a rigid platform, at a distance $c'$ from the bottom of the ocean (see figure \ref{fig:geom}$a,b$). The device is located in the middle of a straight channel, whose impermeable walls are placed at a mutual distance $b'$ and extend to infinity to either side (see figure \ref{fig:geom}$b$). Incident waves are coming from the right with wave crests parallel to the OWSC and set the flap into an oscillating motion. This is eventually converted into useful energy by means of a generator linked to the device. Note that, due to the mirroring effect of the side walls, the layout of figure \ref{fig:geom}($b$) can also be employed to model the behaviour of an infinite array of identical OWSCs in the open ocean, subject to beam seas. The converters in this array have spatial period $b'$ and oscillate at unison. Returning to the channel layout, let us now define a plane reference system of coordinates $\boldsymbol{x}'=(x',y',z')$. Let $x'$ lie on the centre line of the channel, $y'$ along the axis of the OWSC at rest position and $z'$ rise up from the undisturbed water level $z'=0$, positive upwards. The flap is able to oscillate on the vertical plane $(x',z')$ about the horizontal axis at $(x',z')=(0,-h'+c')$, thus representing a system with one degree of freedom, i.e. pitch. Its time-dependent amplitude of rotation $\theta'=\theta'(t')$ is defined positive if counter-clockwise; $t'$ denotes time. We also assume that the bottom foundation does not oppose the flap movement \citep[ideal foundation, see][]{SA96} and that the hinge is frictionless. As in many gravity-wave problems \citep[see for example][]{ME05} the fluid is deemed to be inviscid and incompressible and the flow irrotational. Hence there exists a potential $\Phi'(x',y',z',t')$ for the velocity field $\boldsymbol{v}'=\nabla ' \Phi'$, where $\nabla '(\cdot) = \left\lbrace(\cdot)_{,x'},(\cdot)_{,y'},(\cdot)_{,z'}\right\rbrace $ is the nabla operator. Subscripts with commas denote differentiation with respect to the relevant variable. The complete set of nonlinear equations governing the system is detailed in Appendix \ref{sec:appA}. Since the resonance of the channel sloshing modes is a linear phenomenon \citep{SKM87}, here we shall consider a linearised theory. The first-order analysis presented in this paper constitutes the basis for further investigation of higher-order phenomena occurring in the system (see \S\ref{sec:disc}). Now let $A'$ denote the amplitude scale of the wave field and $g$ the acceleration due to gravity. Then introduce the following non-dimensional variables
\begin{equation}
(x,y,z)=(x',y',z')/b',\; t=\sqrt{\frac{g}{b'}}\,t',\; \zeta=\zeta'/A',\; \Phi=\left(\sqrt{gb'}A'\right)^{-1}\Phi',\; \theta=\theta'/\epsilon,
\label{eq:nondimvar}
\end{equation}
where $\epsilon=A'/b'$ is the physical scale of the angular rotation, and constants
\begin{equation}
(h, a, w)=(h',a',w')/b'.
\label{eq:nondimcon}
\end{equation}
The linear governing equations are derived  in the limit of small-amplitude oscillations of the flap, i.e. $\epsilon\ll 1$, by considering the terms $O(1)$ in the fully nonlinear system (\ref{eq:laplace})--(\ref{eq:flapmot}) scaled according to (\ref{eq:nondimvar}). The potential is governed by the Laplace equation
\begin{equation}
\nabla^2\Phi=0,\quad (x,y,z)\in\Omega
\label{eq:motlin}
\end{equation}
where $\Omega$ is the fluid domain. On the free-surface, the kinematic-dynamic boundary condition reads
\begin{equation}
\Phi_{,tt}+\Phi_{,z}=0,\quad z=0.
\label{eq:bcfreesurflin}
\end{equation}
The no-flux conditions on the solid boundaries require 
\begin{equation}
\Phi_{,y}=0,\quad y=\pm 1/2
\label{eq:bcwallin}
\end{equation}
on the channel lateral walls and
\begin{equation}
\Phi_{,z}=0,\quad z=-h
\label{eq:bcbottlin}
\end{equation}
on the bottom. Allowing only tangential motion along the lateral surfaces of the OWSC yields
\begin{equation}
\Phi_{,x}=-\theta_{,t}(t)\,(z+h-c)\,H(z+h-c),\quad x=\pm a,\; -w/2<y<w/2,
\label{eq:bcflap2lin}
\end{equation}
where $H$ denotes the Heaviside step function. Finally the linearised equation of motion of the flap reads
\begin{equation}
I\theta_{,tt}(t)+C\theta(t)=-\int_{-h+c}^0 \int_{-w/2}^{w/2} \left[\Phi_{,t}(a,y,z,t)-\Phi_{,t}(-a,y,z,t) \right](z+h-c)\,\de y\de z+\mathcal{T}_{\epsilon}(t),
\label{eq:flapmotlin}
\end{equation}
where $C=wa(h-c)^2-S$ is the elastic  moment of inertia due to net buoyancy, $S=S'/(\rho b'^4)$ and $I=I'/(\rho b'^5)$ are respectively the first and second moment of inertia of the flap and $\mathcal{T}_\epsilon=\mathcal{T}'_\epsilon/(\rho g b'^4)$ is the first-order torque exerted by the generator on the flap, all in non-dimensional variables; $\rho$ is water density. Without loss of generality, we shall assume $\mathcal{T}_\epsilon$ to depend on time via $\theta(t)$ and to be partly inertial, partly elastic and partly damping \citep[see][for a similar example]{ME05}
\begin{equation}
\mathcal{T}_\epsilon(t) = -\mu_{pto}\, \theta_{,tt}(t)-C_{pto}\, \theta(t) - \nu_{pto}\,\theta_{,t}(t).
\label{eq:Teps}
\end{equation}
In the latter expression $\mu_{pto}$ and $C_{pto}$ are respectively the inertial and elastic characteristic of the generator, while $\nu_{pto}$ is the energy extraction rate. Now note that the physical scale of the OWSC width is $a'=O(A')$ \citep{AL08}, so that $a=a'/b'=O(\epsilon) \ll 1$. Hence the thickness of the OWSC can be neglected if compared to the reference length scale $b'$ and the hypothesis of thin obstacle \citep{LM01} can be applied to evaluate the potential $\Phi$. As a consequence, $a= 0$ in the boundary condition on the lateral surfaces (\ref{eq:bcflap2lin}) and in the right-hand side of the equation of motion (\ref{eq:flapmotlin}).
\subsection{Solution of the radiation and scattering problems}\label{sec:radsca}
In this section solution is found to the Laplace equation (\ref{eq:motlin}) with linearised boundary conditions (\ref{eq:bcfreesurflin}) on the free-surface, (\ref{eq:bcwallin}) on the channel walls, (\ref{eq:bcbottlin}) on the bottom and (\ref{eq:bcflap2lin}) on the flap lateral surfaces. Once having solved this system, expression (\ref{eq:flapmotlin}) will be employed to determine the motion of the flap. Assume that the flap undergoes simple harmonic oscillations of frequency $\omega=\omega'\sqrt{b'/g}$ about its rest position, so that
\begin{equation}
\theta(t)=\Real\left\lbrace \Theta \,\e^{-\img\omega t}\right\rbrace,
\label{eq:thetatime}
\end{equation}
where $\Real$ is the real part and $\Theta$ the complex amplitude of rotation. Position (\ref{eq:thetatime}) allows to separate the time factor in the time-dependent variables, so that the velocity potential can be rewritten as
\begin{equation}
\Phi(x,y,z,t)=\phi(x,y,z)\,\e^{-\img\omega t},
\label{eq:phifactor}
\end{equation}
where the symbol $\Real$ is omitted for the sake of brevity. Due to the linearity of the problem, the spatial potential $\phi$ is analysed by resorting to the classical decomposition \citep[see for example][]{LM01,ME05}
\begin{equation}
\phi = \phi^R+\phi^S.
\label{eq:phidecomp}
\end{equation} 
In the latter, $\phi^R$ is the solution of the radiation problem, in which the flap is set to oscillate without incoming incident waves, and $\phi^S$ is the solution of the scattering problem, where the flap is held fixed in incoming waves. In turn, the scattering potential can be decomposed into 
\begin{equation}
\phi^S=\phi^I+\phi^D,
\label{eq:phisdecomp}
\end{equation}
 where
\begin{equation}
\phi^I(x,y,z)=-\frac{\img A_0}{\omega\cosh kh}\cosh k(z+h) \e^{-\img kx}
\label{eq:incidentwav}
\end{equation}
is the potential of the incident wave and $\phi^D$ the potential of the diffracted waves \citep[see][]{LM01, ME05}. In (\ref{eq:incidentwav}) $A_0$ and $k$ are respectively the non-dimensional incident wave amplitude and wavenumber, the latter depending on the wave frequency according to the dispersion relation $\omega^2 = k\tanh kh$. In the following, the radiation and scattering problems will be investigated and their results summed up to obtain the total wave field.

The boundary-value problems for the radiation potential $\phi^R$ and the diffraction potential $\phi^D$ can be determined by employing the factorizations (\ref{eq:thetatime})--(\ref{eq:phifactor}) and the decompositions (\ref{eq:phidecomp})--(\ref{eq:phisdecomp}) into the governing equations (\ref{eq:motlin})--(\ref{eq:bcflap2lin}). As a consequence, $\phi^R$ and $\phi^D$ must satisfy the Laplace equation
\begin{equation}
\nabla^2\phir=0,\quad (x,y,z)\in\Omega,
\label{eq:motphir}
\end{equation}
where the shorthand notation $\phir$ denotes either potential, the kinematic-dynamic boundary condition on the free-surface
\begin{equation}
\phir_{,z}-\omega^2\phir=0,\quad z=0,
\label{eq:bcfreesurfphir}
\end{equation}
the no-flux condition on the bottom
\begin{equation}
\phir_z=0,\quad z=-h
\label{eq:bcbottphir}
\end{equation}
and on the channel walls
\begin{equation}
\phir_{,y}=0,\quad y=\pm 1/2
\label{eq:bcwallphir}
\end{equation}
and the kinematic conditions on the lateral surfaces of the flap
\begin{equation}
\left\lbrace \begin{array}{c}
\phi^R_{,x}\\[5pt] \phi^D_{,x}\end{array}\right\rbrace  = \left\lbrace \begin{array}{c}
\img\omega\Theta\,(z+h-c) H(z+h-c)\\ -\phi^I_{,x}\end{array}\right\rbrace,\quad x=\pm 0,\; -w/2<y<w/2.
\label{eq:bcflapphir}
\end{equation}
Finally, $\phir$ must be outgoing at large $|x|$. Separation of variables 
\begin{equation}
\phir(x,y,z)=\sum_{n=0}^{\infty}\varphi_n^{(R,D)}(x,y) Z_n(z), \quad n=0,1,2,\dots,
\label{eq:sepvar}
\end{equation}
yields the well-known vertical eigenmodes \citep[see for example][]{ME05}
\begin{equation}
Z_n(z)=\frac{\sqrt{2}\cosh \kappa_n (z+h)}{\left(h+\omega^{-2}\sinh^2 \kappa_n h\right)^{1/2}},\quad n=0,1,2,\dots,
\label{eq:Zn}
\end{equation}
which satisfy the orthogonality relation
\begin{equation}
\int_{-h}^0 Z_n(z) Z_m(z)\, \de z=\delta_{nm},
\label{eq:orthrel}
\end{equation}
$\delta_{nm}$ being the Kronecker delta, $n,m\in\mathbb{N}$. In $Z_n$ (\ref{eq:Zn})
\begin{equation}
\kappa_0=k,\quad \kappa_n=\img k_n,\quad n=1,2,\dots
\label{eq:kappan}
\end{equation}
are the solutions of the dispersion relationships
\begin{equation}
\omega^2=k\tanh kh, \quad \omega^2=-k_n \tan k_n h,\quad n=1,2,\dots
\label{eq:disprel}
\end{equation}
respectively. Hence an equivalent form of (\ref{eq:Zn}) for $n>0$ is $Z_n=\sqrt{2}\cos k_n(z+h)/(h-\omega^{-2}\sin^2 k_n h)^{1/2}$, i.e. an oscillating function of $z$. By making use of (\ref{eq:sepvar}) and (\ref{eq:Zn}), the Laplace equation (\ref{eq:motphir}) becomes
\begin{equation}
\left(\nabla^2+\kappa_n^2\right)\varphi^{(R,D)}_n=0, \quad (x,y)\in \Sigma,
\label{eq:helmrad}
\end{equation}
governing the plane radiation and diffraction potentials $\varphi^{(R,D)}_n(x,y)$ in the 2D fluid domain $\Sigma$ (see figure \ref{fig:geom}), where $n=0,1,2,\dots$ will be omitted from now on for brevity. Note that (\ref{eq:helmrad}) is an ordinary Helmholtz equation when $n=0$, for which $\kappa_0^2=k^2$, and a modified Helmholtz equation when $n>0$, for which $\kappa_n^2=-k_n^2$. With the same steps as above, the no-flux condition on the channel walls (\ref{eq:bcwallphir}) gives
\begin{equation}
\varphi^{(R,D)}_{n,y}=0,\quad y=\pm 1/2.
\label{eq:bcwallrad}
\end{equation}
Now consider the kinematic conditions on the flap (\ref{eq:bcflapphir}). Using the factorization (\ref{eq:sepvar}), multiplying both sides by $Z_m$, integrating along the vertical line from $z=-h$ to $z=0$ and using the orthogonality relation (\ref{eq:orthrel}) yields
\begin{equation}
\left\lbrace \begin{array}{c}
\varphi^R_{n,x} \\[5pt] \varphi^D_{n,x}\end{array}\right\rbrace=\left\lbrace \begin{array}{c}
V f_n \\ A_0\,d_n\end{array}\right\rbrace ,\quad x=\pm 0,\; -w/2<y<w/2.
\label{eq:bcflaprad}
\end{equation}
In the latter expression
\begin{equation}
V=\img\omega\Theta
\label{eq:V}
\end{equation}
is the complex angular velocity of the flap, while
\begin{equation}
f_n=\frac{\sqrt{2}\left[\kappa_n (h-c)\sinh \kappa_n h+\cosh \kappa_n c-\cosh \kappa_n h\right]}{\kappa_n^2\left(h+\omega^{-2}\sinh^2\kappa_n h\right)^{1/2}}
\label{eq:Fn}
\end{equation}
and
\begin{equation}
d_0=\frac{k\left(h+\omega^{-2} \sinh^2 kh\right)^{1/2}}{\sqrt{2}\,\omega\cosh kh},\quad d_n=0,\quad n=1,2,\dots
\label{eq:Dn}
\end{equation}
are real constants depending on the modal order $n$. Finally $\varphi^{(R,D)}_n$ must be outgoing disturbances along the channel. In summary, the 2D $n$-th mode potentials $\varphi^{(R,D)}_n$ must solve the Helmholtz equation (\ref{eq:helmrad}) with the boundary conditions (\ref{eq:bcwallrad})--(\ref{eq:bcflaprad}) and be outgoing. Here we shall solve this boundary-value problem by employing an integral-equation method based on the application of Green's theorem with an appropriate channel Green function. In the literature \citet{AL86} and \citet{MR89} already used a similar procedure to solve plane sound-wave problems, while \citet{PM92,PM94,PM95} applied this method to the scattering and trapping of surface waves by plates. Later, \citet{MF97} used the integral-equation approach to solve the radiation problem for a heaving submerged horizontal disc, but without bounding walls and in water of infinite depth. In this paper, the boundary-value problem (\ref{eq:helmrad})--(\ref{eq:bcflaprad}) is solved in Appendix \ref{sec:appB} by applying an integral-equation technique based on the decomposition of the Green function into a singular part and an analytical remainder. Such a procedure is inspired by \citet{CH94}'s analysis of the tank Green function for numerical models. The integral-equation technique allows to solve the plane boundary-value problems for $\varphi^{(R,D)}_n$ and ultimately to express the potentials $\phir$ (\ref{eq:sepvar}) in an appealing fast-converging semi-analytical form (see Appendix \ref{sec:appB} for details). The radiation potential is given by
\begin{eqnarray}
\phi^R(x,y,z)=-\frac{\img wV}{8}\sum_{n=0}^\infty \kappa_n x\, Z_n(z)\sum_{p=0}^{M}\alpha_{(2p)n}\sum_{m=-\infty}^{+\infty}\int_{-1}^1 \left(1-u^2\right)^{1/2} U_{2p}(u)\nonumber\\
\times\frac{\Haa\left({\kappa_n\sqrt{x^2+(y-\shalf wu-m)^2}}\right)}{\sqrt{x^2+(y-\shalf wu-m)^2}}\,\de u,
\label{eq:radpotsimp}
\end{eqnarray}
where $\Haa$ is the Hankel function of the first kind and first order and the $U_{2p}$ are the Chebyshev polynomials of the second kind and order $2p, p=0,1\dots M\in \mathbb{N}$. Finally, the $\alpha_{(2p)n}$ are the complex solutions of the linear system of equations (\ref{eq:linsysrad}) which ensures that the boundary condition on the plate (\ref{eq:bcflaprad}) is satisfied. This system is solved numerically with a collocation scheme (see again Appendix \ref{sec:appB}), therefore the solution is partly numerical. Note that the radiation potential (\ref{eq:radpotsimp}) is an odd function of $x$, as suggested by the antisymmetric pitching motion of the flap. Physically, (\ref{eq:radpotsimp}) represents the sum of outgoing waves within the waveguides imposed by the lateral walls and is similar in form to the solutions of other problems of water waves propagating past sharp obstacles \citep[see for example][who also employ Chebyshev polynomials]{EP97,L10}. The inner series in (\ref{eq:radpotsimp}) represents a sum of disturbances radiated by sources arranged periodically along the $y$ axis and converges as $m^{-3/2}$ for large $m$. Numerically, summation from $m=-20$ up to $m=20$ ensures convergence of the series with relative error $O(10^{-3})$ close to resonance and $O(10^{-4})$ elsewhere. The central series in (\ref{eq:radpotsimp}) is truncated at $M<\infty$ and its coefficients are calculated numerically as described in Appendix \ref{sec:appB}. Numerical investigation showed that $M\geq 4$ is sufficient to ensure convergence with virtually no relative error. The outer series in (\ref{eq:radpotsimp}) over the vertical eigenmodes was found to converge quickly already for $n=4$ with a relative error $O(10^{-5})$. Hence the method of solution adopted here, based on the appropriate decomposition of the Green function (\ref{eq:Gseries})--(\ref{eq:Gn1}), proves to be reliable and computationally efficient. Turning to the scattering problem, the diffraction potential is given by
\begin{eqnarray}
\phi^D(x,y,z)=-\frac{\img w A_0}{8}\,kx Z_0(z)\sum_{p=0}^{M}\beta_{(2p)0}\sum_{m=-\infty}^{+\infty}\int_{-1}^{1} \left(1-u^2\right)^{1/2} U_{2p}(u)\nonumber\\
\times\frac{\Haa\left(k\sqrt{x^2+(y-\shalf wu-m)^2} \right)}{\sqrt{x^2+(y-\shalf wu-m)^2}}\,\de u.
\label{eq:difpot}
\end{eqnarray}
In the latter expression the $\beta_{(2p)0}$ are the complex solutions of the linear system of equations (\ref{eq:linsysrad}) for $n=0$, which ensures that the no-flux condition on the flap (\ref{eq:bcflapphir}) is satisfied. Thanks to linearity, the total potential $\phi=\phi^R+\phi^D+\phi^I$ is obtained by simply summing up the solutions  $\phi^R$ (\ref{eq:radpotsimp}) and $\phi^D$  (\ref{eq:difpot}) of the radiation and diffraction problem, respectively, to the incident wave potential $\phi^I$ (\ref{eq:incidentwav}). The equation of motion of the flap can now be solved.
\subsection{Flap motion}
\subsubsection{Hydrodynamic parameters}\label{sec:hydro}
Consider the law of motion of the flap (\ref{eq:flapmotlin}). By using the factorizations (\ref{eq:thetatime}) and (\ref{eq:phifactor}) for $\theta$ and $\Phi$ respectively, applying the decomposition $\phi=\phi^R+\phi^D+\phi^I$, employing the solutions (\ref{eq:radpotsimp}) and (\ref{eq:difpot}) for $\phi^R$ and $\phi^D$ respectively and expressing $\mathcal{T}_\epsilon$ with (\ref{eq:Teps}), (\ref{eq:flapmotlin}) yields
\begin{equation}
\left[-\omega^2 (I+\mu+\mu_{pto})+C+C_{pto}-\img\omega(\nu+\nu_{pto})\right]\Theta=F.
\label{eq:flapmotfin}
\end{equation}
In the latter expression $\Theta$ is the unknown complex amplitude of rotation of the flap, while
\begin{equation}
\mu =\frac{\upi w}{4}\,\Real\left\lbrace \sum_{n=0}^{+\infty}\alpha_{0n}f_n \right\rbrace
\label{eq:addmass}
\end{equation}
is the added torque due to inertia and
\begin{equation}
\nu = \frac{\omega \upi w}{4}\,\Imag\left\lbrace \sum_{n=0}^{+\infty}\alpha_{0n}f_n \right\rbrace 
\label{eq:raddamp}
\end{equation}
is the radiation damping, both depending on the solutions $\alpha_{0n}$ to the linear system (\ref{eq:linsysrad}) for the radiation problem. Finally
\begin{equation}
F=-\frac{\upi w}{4}\,\img\omega A_0 \beta_{00} f_0
\label{eq:Td}
\end{equation}
is the complex excitation torque, corresponding to the action on the plate as if it was held fixed in incoming waves.
\subsubsection{Extracted power and capture factor}
Expression (\ref{eq:flapmotfin}) formally describes a simple harmonic oscillator in the frequency domain; hence the average extracted power from the OWSC over a period $T=2\upi/\omega$ is
\begin{equation}
P=\frac{1}{T}\int_0^T \left(\nu_{pto}\,\theta_{,t}\right)\,\theta_{,t}\,\de t=\frac{1}{2}\frac{\omega^2\nu_{pto}\left|F \right|^2}{\left[C+C_{pto}-\left(I+\mu+\mu_{pto} \right)\omega^2 \right]^2+\left(\nu+\nu_{pto}\right)^2\omega^2},
\label{eq:Pto}
\end{equation}
where again $\theta=\Real\left\lbrace\Theta \e^{-\img\omega t}\right\rbrace$ and $\Theta$ is the solution of (\ref{eq:flapmotfin}).  Returning to physical variables via (\ref{eq:nondimvar}), the power take-off expressed in Watts $(\mathrm{W})$ is then
\begin{equation}
P'=\rho A'^2b'^{3/2}g^{3/2} P \quad\left[\mathrm{W}\right].
\label{eq:Ptop}
\end{equation}
The rate of power take-off (\ref{eq:Pto}) can be maximized according to the optimizing criteria
\refstepcounter{equation}
$$
  \omega=\omega_0=\sqrt{\frac{C+C_{pto}}{I+\mu+\mu_{pto}}}\,,\quad \nu_{pto}=\nu.
  \eqno{(\theequation{\mathit{a},\mathit{b}})}\label{eq:Poptimize}
$$
In the latter, condition (\ref{eq:Poptimize}$a$) corresponds to the tuning between the incident wave frequency $\omega$ and the natural pitching frequency $\omega_0$ of the flap, while (\ref{eq:Poptimize}$b$) requires the equality of the rate of power take-off $\nu_{pto}$ to the radiation damping $\nu$ (\ref{eq:raddamp}).
A quick numerical assessment of the tuning condition (\ref{eq:Poptimize}$a$) for a typical OWSC configuration (see \S\ref{sec:expcomp}) reveals that the natural pitching period $T_0'=2\upi/\omega_0\sqrt{b'/g}\simeq 20\,\mathrm{s}$ is larger than the ordinary wave periods on site, about $5-15\,\mathrm{s}$ \citep{AL08}. Hence the tuning condition (\ref{eq:Poptimize}$a$) is not achieved in normal operating circumstances. 
Away from body resonance, the extracted power (\ref{eq:Pto}) can still be optimized by adjusting the damping rate $\nu_{pto}$ such as $\partial P/\partial \nu_{pto}=0$, which yields 
\begin{equation}
\nu_{pto}=\nu_{opt}=\sqrt{\frac{\left[C+C_{pto}-(I+\mu+\mu_{pto})\omega^2 \right]^2}{\omega^2}+\nu^2}
\label{eq:nuopt}
\end{equation}
as the optimum damping rate \citep[see][]{FA02}. Then the optimum power $P_{opt}$ is obtained by substituting $\nu_{pto}=\nu_{opt}$ in (\ref{eq:Pto}). 
Now, the rate of extracted power and the optimum power are functions of the wave climate, so that the amount of energy extracted from a given floating body strongly depends on the sea state. Therefore both the rate of power take-off and the optimum power are not reliable indicators of the efficiency of the system \citep{CR08}. A more rational efficiency assessment is done by considering the capture factor, defined as the ratio between the optimum power extracted per unit flap width and the power of the incident wave field per unit crest length, i.e.
\begin{equation}
C_F = \frac{P_{opt}'}{\shalf\rho g C'_g A_0'^{\,2} w'},
\label{eq:capwid}
\end{equation}
where $A_0'=A_0A'$ and
\begin{equation}
C_g'=\frac{\omega'}{2k'}\left(1+\frac{2k'h'}{\sinh 2k'h'}\right)\quad \left[\mathrm{m}/\mathrm{s}\right]
\label{eq:Cgp}
\end{equation}
are respectively the amplitude and the group velocity of the incident waves in physical variables. In the following, all the parameters describing the performance of the OWSC will be assessed in physical variables.

In the next sections, the analytical theory will be validated against numerical and experimental models. Discussion will follow on the physical behaviour of the system and further analysis will investigate the dependence of the system on its main parameters, such as the frequency of the incident waves and the width of the flap. Resonance will be shown to occur at special frequencies, namely the cut-off frequencies of sloshing waves in open channels. The response curves of the system dynamic characteristics show that the OWSC performance is largest under resonance conditions.
\section{Discussion}\label{sec:disc}
\subsection{Numerical and experimental comparison}\label{sec:expcomp}
\begin{figure}
  \centerline{\includegraphics[width=14cm, trim= 0cm 0cm 0.5cm 0cm]{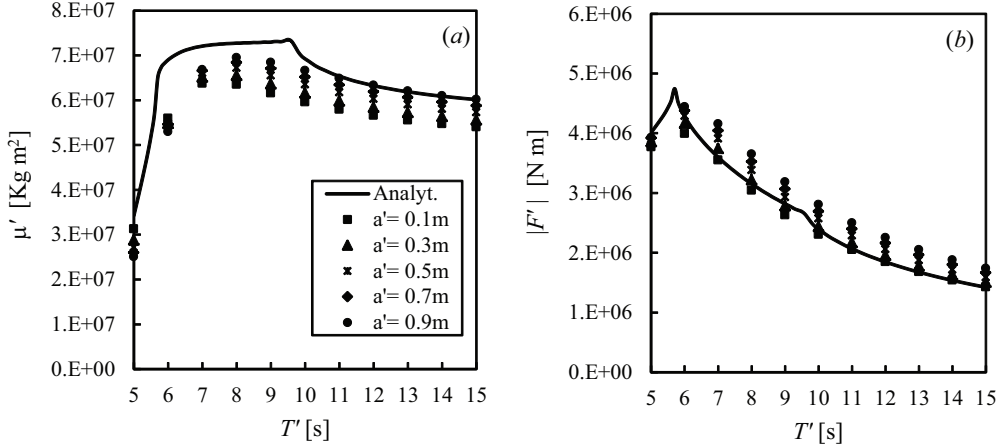}}
  \caption{Behaviour of (\textit{a}) added inertia torque and (\textit{b}) excitation torque with wave period in physical variables. The solid line refers to the analytical values of \S\ref{sec:hydro}, while the markers refer to the numerical solution of \citet{JO09} for various thicknesses of the flap.}
\label{fig:numcomp}
\end{figure}

\begin{figure}
  \centerline{\includegraphics[width=7.5cm]{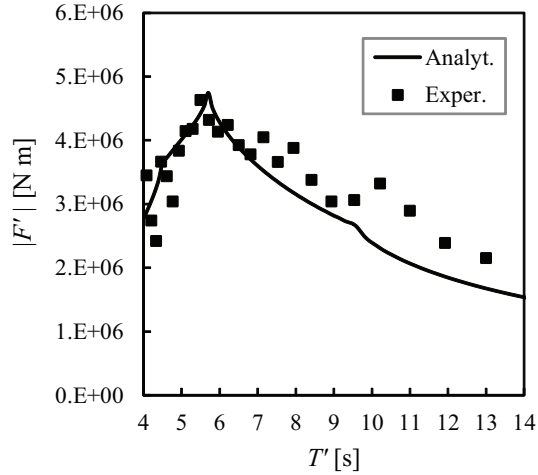}}
  \caption{Behaviour of the excitation torque with the period of the incident waves in physical variables. The squares refer to the experimental results of \citet{AL08} for the $20^{\mathrm{th}}$ scale model of the Oyster$^\mathrm{TM}$ WEC, while the solid line refers to the equivalent analytical model as developed in section \S\ref{sec:anmod}. Discrepancy between data at $T'\simeq$ 10 s can be explained by the end-to-end resonance of long-crested waves in the wave tank.}
\label{fig:expcomp}
\end{figure}

In order to validate the theory, comparison is made with the numerical results of \citet{JO09}. In the analytical model, the width of the flap is $w'=18\,\mathrm{m}$, while the width of the channel is $b'=91.6\,\mathrm{m}$, giving a blockage ratio $w=w'/b'\simeq0.2$. The depth of the channel is $h'=10.9\,\mathrm{m}$ and the distance between the bottom and the hinge is $c'=1.5\,\mathrm{m}$. The amplitude of the incident wave is $A'_0=0.3\,\mathrm{m}$, which gives $\epsilon=A'_0/b'\simeq 0.003\ll 1$. This justifies the assumption of linear theory for this case.
The numerical data have been obtained with the software package WAMIT \citep{WAM} for the linear analysis of the interaction of surface waves with offshore structures. The geometry of the flap used in the numerical simulations is that of an $18\,\mathrm{m}$ wide rectangular box standing on a triangular prism pointing downwards \citep[see][for specifications]{JO09}. The flap in turn is hinged on a $1.5\,\mathrm{m}$ high rectangular platform lying on the bottom of a $10.9\,\mathrm{m}$ deep ocean. In the numerical calculations of \citet{JO09} the device is placed in open water, so that the model does not simulate the effect of the lateral walls. Figures \ref{fig:numcomp}($a$) and \ref{fig:numcomp}($b$), respectively, show the values of the added inertia torque $\mu'=\rho b'^{5}\mu$ and the absolute value of the excitation torque $\left|F'\right|=\rho g A'b'^{3}|F|$  in physical variables versus the wave period $T'$, for both the analytical and the numerical model. 
While in the analytical model the flap has zero thickness, in the numerical model the simulations have been carried on with different non-zero values of the total thickness $2a'$ (see again figure \ref{fig:numcomp}$a$,$b$). The agreement between both models is satisfactory. However for the added inertia torque $\mu'$ (figure  \ref{fig:numcomp}$a$), the analytical values are slightly larger than the numerical ones. This is likely due to the differences in the shape of the flap, a perfect rectangular box in the analytical model and a rectangular box sitting on a triangular base in the numerical model. Also, note that the analytical values are larger than the numerical ones next to the peaks of either the added inertia torque ($T'\simeq 9.6\,\mathrm{s}$ in figure \ref{fig:numcomp}$a$) and the excitation torque ($T'\simeq 5.7\,\mathrm{s}$ in figure \ref{fig:numcomp}$b$). This is a consequence of the bounding effect of the lateral walls, absent in the numerical model of \citet{JO09}, that will be discussed in depth in \S\ref{sec:resonance}. Finally, the numerical results in figure \ref{fig:numcomp}($a$,$b$) clearly show that the hydrodynamic coefficients do not vary noticeably while varying the flap thickness. The initial assumption of neglecting the thickness of the flap to calculate the hydrodynamic actions, namely the thin-plate approximation, is then validated. The theoretical results of \S\ref{sec:anmod} are further validated by comparison with the data of \citet{AL08}. These have been obtained during an experimental campaign at Queen's University Belfast on a $20^{\mathrm{th}}$ scale model of the Oyster$^\mathrm{TM}$ wave energy converter developed by Aquamarine Power Ltd \footnote{www.aquamarinepower.com} \citep[see][]{FWJ07, Wetal07, Hetal10} having similar dimensions to those already specified for the theoretical models. In figure \ref{fig:expcomp} the behaviour of the excitation torque $\left|F'\right|$ with the incident wave period $T'$ is reported for both the equivalent analytical box-shaped flap and the experimental model of the Oyster$^\mathrm{TM}$ WEC.
Very good agreement is found between the two sets of data: the theoretical model described in \S\ref{sec:anmod} is also capable to simulate the behaviour of a real - and more complex - OWSC.
\subsection{Resonance}\label{sec:resonance}
\begin{figure}
  \centerline{\includegraphics[width=7.5cm]{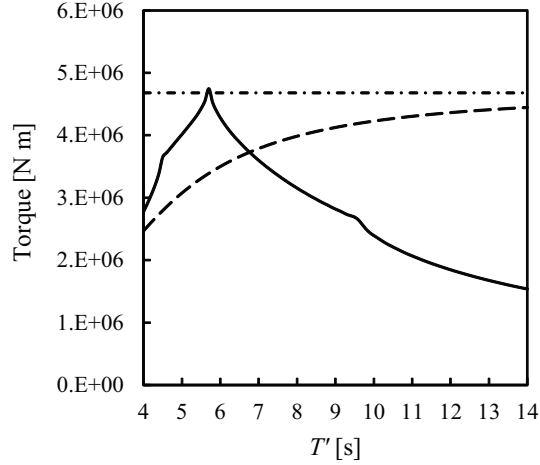}}
  \caption{Behaviour of the excitation torque with the period of the incident waves in physical variables. The solid line refers to the 3D value $F'$, calculated for the geometry of \S\ref{sec:expcomp} . The dashed line refers to the 2D value $F_{2d}'$ (\ref{eq:Td2d}) for the simplified 2D channel, where $b'=w'=18\,\mathrm{m}$. The dash-dotted line represents the 2D long-wave limit (\ref{eq:2dlwlim}). Note that the 3D torque has a maximum at $T'=5.7\,\mathrm{s}$ and spikes at $T'= 4.5\,\mathrm{s}$ and $T'= 9.6\,\mathrm{s}$.}
\label{fig:modcomp}
\end{figure}

\begin{figure}
  \centerline{\includegraphics[width=8.5cm]{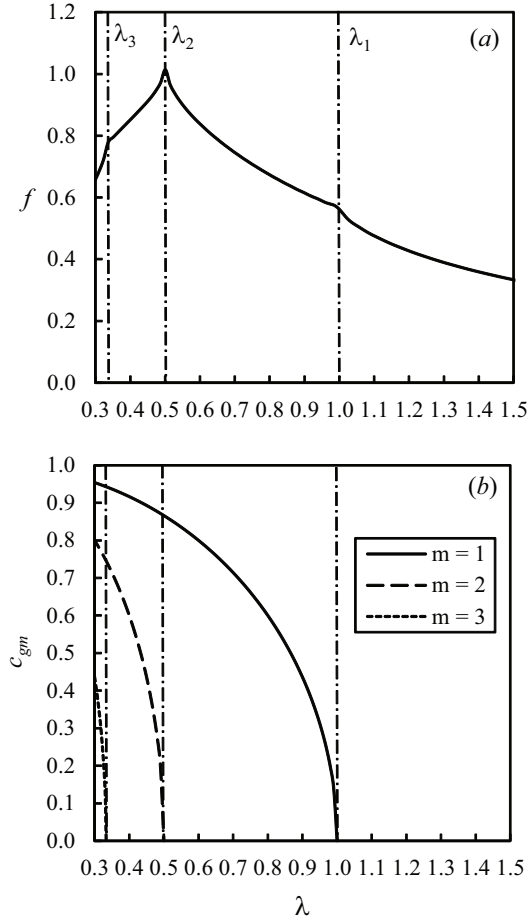}}
  \caption{($a$) Non-dimensional excitation torque ratio $f$ (\ref{eq:fd}) versus the wavelength of the incident wave for the geometry of \S\ref{sec:expcomp} . Note the spiky shape of the curves, with spikes occurring at the resonant frequencies $\lambda_1=1$, $\lambda_2=0.5$ and $\lambda_3=0.333\dots$ marked by vertical dash-dotted lines. ($b$) Group velocity ratio $c_{gm}$ (\ref{eq:cgm}) versus wavelength for the first three sloshing modes. Note that the group velocity of each sloshing mode decays as soon as the relevant cut-off wavelength $\lambda_m$ (\ref{eq:lambdam}) is approached.}
\label{fig:sloshmod}
\end{figure}

\begin{figure}
  \centerline{\includegraphics[width=13 cm, trim= 0cm 0cm 0.5cm 0cm]{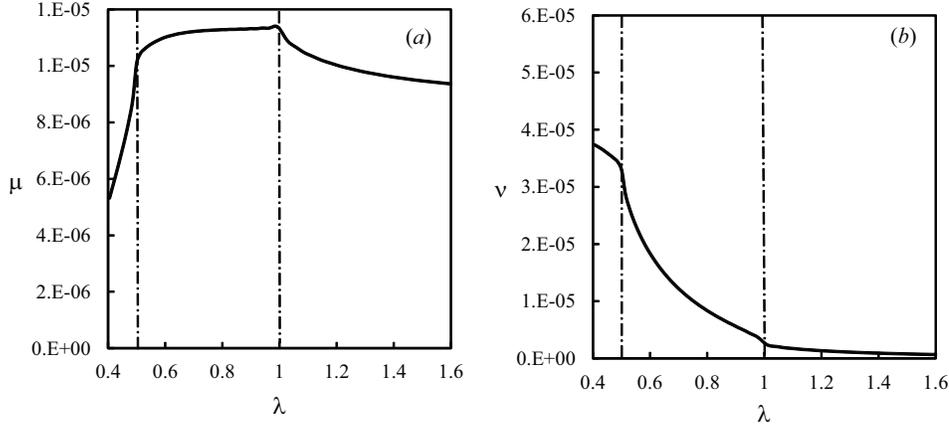}}
  \caption{Behaviour of (\textit{a}) added inertia torque and (\textit{b}) radiation damping with the wavelength of the generated waves in non-dimensional variables for the geometry of \S\ref{sec:expcomp}. Vertical dash-dotted lines indicate the resonant frequencies of the transverse sloshing modes.}
\label{fig:radplot}
\end{figure}

After having shown a good agreement between the analytical and the numerical/exper- imental data, we shall investigate in more depth the 3D dynamics of the flap in the channel and the influence of the channel lateral walls in enhancing the efficiency of the converter. First, let us consider the diffraction problem and analyse the modifications induced by the 3D wave field in the 3D channel model of \S\ref{sec:anmod} with respect to a simplified reference model, in which $b'=w'$, i.e. the flap spans the entire width of the channel. Here the behaviour of the fluid is clearly two dimensional, so that this layout will be referred to as the 2D channel. As already  demonstrated in the literature \citep[]{LM01, FA02}, the excitation torque acting on the flap in the 2D channel is, in physical variables,
\begin{equation}
F_{2d}' = 2\rho g A_0' w' \left[\frac{h'-c'}{k'}\,\tanh k'h' + \frac{\cosh k'c' - \cosh k'h'}{k'^2 \cosh k'h'}\right] \quad \left[ \mathrm{Nm}\right].
\label{eq:Td2d}
\end{equation}  
In figure \ref{fig:modcomp} the excitation torque is plotted against the wave period $T'$ in physical variables, for either the 3D-channel model of \S\ref{sec:anmod} and the 2D-channel reference model.
While the 2D excitation torque regularly increases with the period up to the asymptotic long-wave value 
\begin{equation}
\bar{F}_{2d}'=\lim_{k\rightarrow 0} F_{2d}'=\rho g w' A_0' (h'-c')^2\; [\mathrm{Nm}]\simeq 4.7\cdot10^6\, \mathrm{Nm},
\label{eq:2dlwlim}
\end{equation}
the 3D torque $|{F}'|$ reaches its maximum value $\max\left\lbrace \left|{F}'\right|\right\rbrace\simeq 4.75\cdot10^6\,\mathrm{Nm}$ at a finite period $T'= 5.7\,\mathrm{s}$. Then it decreases while moving towards longer periods, with a secondary spike at $T'= 9.6\,\mathrm{s}$. Hence an immediately noticeable effect of the 3D dynamics is the lowering of the peak period to finite values. Though the two curves of figure \ref{fig:modcomp} have a very different aspect, they show the interesting property that $\max\left\lbrace {|F|}'\right\rbrace / \max\left\lbrace  {F}'_{2d} \right\rbrace=O(1)$ (see again figure \ref{fig:modcomp}). Hence to further study the 3D effects on the system, it is convenient to consider the non-dimensional ratio
\begin{equation}
f=\frac{\left| {F}'\right|}{ \bar{F}'_{2d} }
\label{eq:fd}
\end{equation}
between the excitation torque in the 3D channel and the maximum excitation torque in the 2D channel. In figure \ref{fig:sloshmod}($a$) $f$ is plotted versus the non-dimensional wavelength $\lambda=\lambda'/b'$ of the incident wave, for the same geometry as above. 
The graph of figure \ref{fig:sloshmod}($a$) exhibits the typical spiky behaviour already noticed by \cite{AL86} when studying the reflection coefficients for normal incidence of sound waves on an array of screens.  Similar spikes have also been found by \citet{ET85}, \citet{YS89}, \citet{LE92} and \citet{CH94} in the analysis of the hydrodynamic coefficients of bodies of various shapes in a channel. Note that in figure \ref{fig:sloshmod}($a$) the spikes, either in the form of local maxima or changes in slope, occur at special values of $\lambda$, i.e. $\lambda = 1,1/2,1/3,\dots$
The latter correspond to the well-known values of the cut-off wavelengths of the symmetric sloshing modes in a channel \citep[see for example][]{SKM87, ME05}
\begin{equation}
\lambda_m=\frac{\lambda'_m}{b'}=\frac{1}{m},\quad m=1,2,\dots
\label{eq:lambdam}
\end{equation}
The sloshing modes are transverse standing waves whose crests are parallel to the axis of the channel, oscillating back and forth between the two bounding lateral walls. The physical behaviour of the sloshing waves is strongly dependent on the width of the channel and on the wavelength of the incident wave. As shown by \citet{ME05}, if $\lambda'/b'=\lambda<\lambda_m$, then the $m$-th sloshing component propagates along the channel, carrying the associated wave energy towards infinity. On the other hand, if $\lambda>\lambda_m$, the amplitude of the $m$-th sloshing mode decays exponentially while moving along the channel. In  our case, this means that the energy associated to these modes is trapped near the OWSC and available for extraction. A further insight on this fundamental influence of the sloshing modes can be achieved by considering the ratio between the $m$-th sloshing mode group velocity $C_{gm}'=C_g'  \sqrt{1-(m\lambda)^2}$ \citep[see][]{ME05} and the group velocity of the incoming waves $C_g'$ (\ref{eq:Cgp}), i.e.
\begin{equation}
c_{gm}=C_{gm}'/C_g'=\sqrt{1-(m\lambda)^2}.
\label{eq:cgm}
\end{equation}
Recall that the group velocity is the speed of energy transport. As soon as $\lambda\rightarrow\lambda_m=1/m$, $c_{gm}\rightarrow 0$ and the energy of the $m$-th sloshing mode is no longer propagating along the channel: the sloshing mode is trapped near the flap. Figure \ref{fig:sloshmod}($b$) shows the behaviour of the group velocity ratio $c_{gm}$ (\ref{eq:cgm}) for the first three sloshing modes versus the wavelength of the incident wave.  Let us first consider the group velocity ratio for the mode $m=3$ (dotted line). When $\lambda\rightarrow\lambda_3=1/3$, the group velocity ratio $c_{g3}$ quickly decays to zero, so that the third sloshing mode is now trapped near the flap. The trapping of the energy associated to this mode determines a sudden increase of the excitation torque acting on the flap, which results in $f$ having a spike at $\lambda=1/3$, as shown in figure \ref{fig:sloshmod}($a$). By further increasing $\lambda$, the group velocity ratio of the second sloshing mode quickly decays too, being $c_{g2}=0$ for $\lambda=\lambda_2=1/2$ (dashed line of figure \ref{fig:sloshmod}$b$). Due to the trapping of the second sloshing mode, the excitation torque acting on the flap further increases, leading to the peak of $f$ occurring exactly at the cut-off frequency of the second sloshing mode, $\lambda=\lambda_2$ (see figure \ref{fig:sloshmod}$a$). Additionally increasing $\lambda$ makes the incoming waves so long that they pass through the flap almost as a uniform swelling. As a consequence, the net pressure acting on the flap diminishes, leading to a decrease of the excitation torque $F$ which also reflects on $f$ (see again figure \ref{fig:sloshmod}$a$). Nevertheless, as soon as $\lambda\rightarrow\lambda_1=1$, $c_{g1}\rightarrow 0$ (see the solid line in figure \ref{fig:sloshmod}$b$) and the first sloshing mode is also trapped near the flap. This results in $f$ having a ultimate spike at $\lambda=1$, as shown in figure \ref{fig:sloshmod}($a$). The physical picture is now clear. The incoming plane waves, initially 2D, impact the flap in the middle of the channel. Differently from the 2D channel, in which the incident waves are totally reflected back to the source, in the 3D channel the waves are also transmitted beyond the flap, due to the lateral gaps between the body and the channel walls. Diffraction at the edges of the flap modifies the behaviour of the wave field and dominates the shape of the hydrodynamic coefficient curves, making them different from the 2D scenario (see figure \ref{fig:modcomp}). As a consequence of this enriched dynamics, transverse sloshing modes, a feature of the channel, are ultimately excited. 
Hence the initial 2D motion shatters into a series of symmetric 3D sloshing waves, resonating each at its own cut-off frequency. When a sloshing mode resonates, the relevant energy is trapped near the flap, resulting in an increase of the excitation torque. Similar conclusions can be drawn also for the radiation problem, where the waves are generated by the pitching motion of the plate. Figure \ref{fig:radplot}($a,b$) shows the plots of the added inertia torque $\mu$ (\ref{eq:addmass}) and the radiation damping $\nu$ (\ref{eq:raddamp}) versus the wavelength of the generated waves in non-dimensional variables. Note that spikes occur here at the same resonant wavelengths as those of $f$ (\ref{eq:fd}) for the scattering problem (see again figure \ref{fig:sloshmod}). This correspondence is due to the relationship between the radiation and scattering problems
$$
F=-\frac{\img\upi w}{8}A_0\frac{\sqrt{k}\left(2kh+\sinh 2kh\right)^{1/2}}{\cosh kh}\,\alpha_{00}
$$
shown in Appendix \ref{sec:appC}, for which the excitation torque acting on the fixed flap in incoming waves is associated to the radiation potential of the pitching motion of the body. We now investigate the OWSC performance.
\subsection{Wave power extraction}\label{sec:power}
\begin{figure}
  \centerline{\includegraphics[width=12.5cm, trim= 0cm 0cm 1cm 0cm]{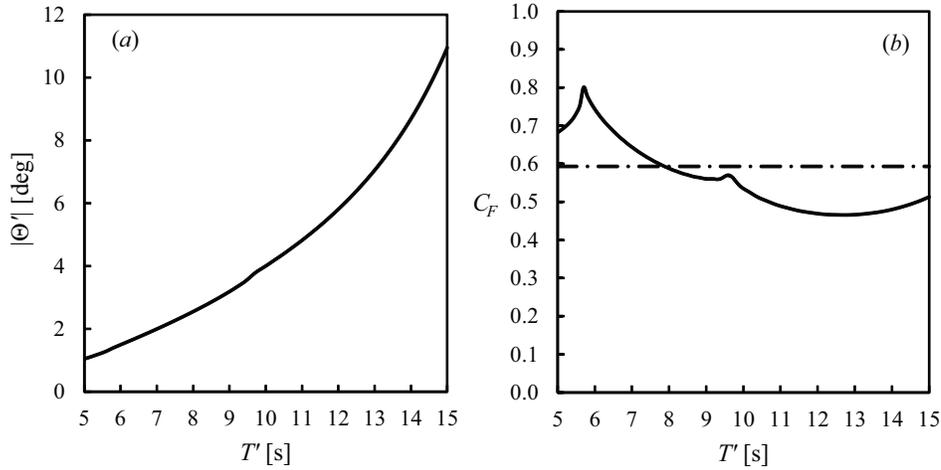}}
  \caption{Behaviour of (\textit{a}) amplitude of rotation and (\textit{b}) capture factor with the period of the incident wave in physical variables for the geometry of \S\ref{sec:expcomp}. In panel ($b$) the horizontal dash-dotted line indicates the average capture factor.}
\label{fig:bodymot}
\end{figure}

After having analysed the radiation and scattering problems separately, we now turn to study the performance of the OWSC in terms of wave power extraction. First, let us consider the flap pitching amplitude $\Theta$, solution of (\ref{eq:flapmotfin}). For easiness of representation, but without loss of generality, in expression (\ref{eq:flapmotfin}) we take the generator inertial and linear terms $\mu_{pto}=C_{pto}=0$ and make use of the optimizing criterion (\ref{eq:nuopt}) for the rate of power take-off, such that $\nu_{pto}=\nu_{opt}$. Finally, we use the values of the moment of inertia $I$ and buoyancy factor $C$ for a typical configuration of the Oyster$^\mathrm{TM}$ wave power device (private communication with Aquamarine Power Ltd). In figure \ref{fig:bodymot}($a$) the absolute value of the flap pitching amplitude is plotted versus the period of the incident wave in physical variables. 
Note that the amplitude of oscillation of the flap increases smoothly while moving to longer waves, not showing the characteristic spiky behaviour of the hydrodynamic coefficients. Since $\Theta$ is related to the displacement in the $x$ direction via the pitching movement of the flap, it seems unaffected by the sloshing modes resonating in the orthogonal $y$ direction.
Let us now consider the capture factor $C_F$, defined by (\ref{eq:capwid}) as a measure of the OWSC efficiency in catching the energy available in incident waves. In figure \ref{fig:bodymot}($b$) the capture factor is plotted against the incident wave period. Comparison between figures \ref{fig:modcomp} and \ref{fig:bodymot}($b$) reveals that both the excitation torque and the capture factor have a peak at $T'=5.7\,\mathrm{s}$, which corresponds to the resonant period of the second sloshing mode. Hence the capture factor is dominated by the excitation torque, which is the power generating action. Overall, figure \ref{fig:bodymot}($b$) reveals that the average capture factor of the 3D-channel model is larger than the maximum theoretical value $C_F=\shalf$ of the 2D-channel model \citep[see for example][]{ME05}. This shows again that the 3D dynamics described in \S\ref{sec:resonance} altogether increases the efficiency of the device with respect to the 2D layout. Finally, the maximum capture factor at resonance is $\max C_F\simeq 0.80$, indicating that the interference between the transverse sloshing modes occurring in the 3D channel can further enhance the OWSC performance.  Resonance of the sloshing modes is therefore beneficial for wave energy extraction. Despite the fact that the dynamics of the OWSC in the channel is affected by the bounding effect of the lateral walls, the general behaviour of the device in the open ocean can also be inferred. There the dominant physical mechanism is still the transmission of the incident waves beyond the flap, but no sloshing modes are present. Hence the OWSC is expected to perform in the open ocean similarly as it does in the 3D channel, exception made for the narrow bandwidths close to the resonant frequencies of the sloshing modes. However, to simulate properly the behaviour of a single device in the open ocean a different mathematical model is needed, in which the waves are allowed to propagate in all directions. In summary, care should be taken when using experimental results obtained in wave tank testing as a benchmark for predicting real on-site behaviour.    
In the following we shall investigate the influence of the system main parameter, namely the width ratio $w=w'/b'$, on the performance of the OWSC.
\subsection{Parametric analysis and further research directions}
\begin{figure}
  \centerline{\includegraphics[width=7cm]{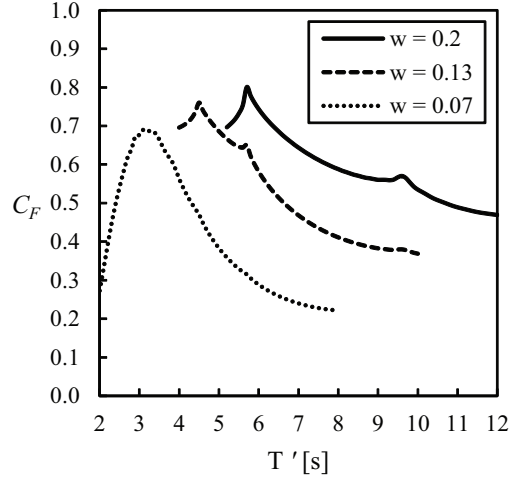}}
  \caption{Behaviour of the capture factor $C_F$ (\ref{eq:capwid}) versus the wave period $T'$ for various flap widths. The solid line refers to $w=0.2$ ($w'=18\,\mathrm{m}$ in physical variables), the dashed line to $w=0.13$ ($w'=12\,\mathrm{m}$) and the dotted line to $w=0.07$ ($w'=6\,\mathrm{m}$).}
\label{fig:cf}
\end{figure}

\begin{figure}
  \centerline{\includegraphics[width=7cm]{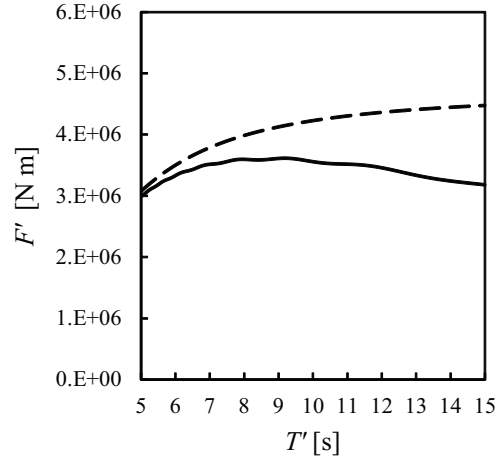}}
  \caption{Behaviour of the excitation torque with the wave period in physical variables. The solid line shows the results for the 3D model with $b'=18\,\mathrm{m}$ and $w'=17.8\,\mathrm{m}$, i.e. $w=0.99$. The dashed line shows the results for the 2D model (\ref{eq:Td2d}) where $b'=w'=18\,\mathrm{m}$, i.e. $w=1$.}
\label{fig:2dlim}
\end{figure}

The response of the capture factor $C_F$ (\ref{eq:capwid}) to variations of the flap width is shown in figure \ref{fig:cf}. Here the plots of $C_F$ against the wave period $T'$ are shown for different widths of the OWSC, i.e. $w'=6\,\mathrm{m}$ (corresponding to $w=w'/b'=0.07$), $w'=12\,\mathrm{m}$ ($w=0.13$) and $w'=18\,\mathrm{m}$ ($w=0.2$), for fixed width of the channel, $b'=91.6\,\mathrm{m}$.
Note that for all widths the capture factor is largest in short periods, thus matching the experimental observations of \citet{Wetal07}. Indeed at larger periods (not shown here), tuning of the incoming wave frequency with the natural pitching frequency of the flaps (\ref{eq:Poptimize}$a$) can occur, thus yielding additional peaks in the capture factor curves. Unfortunately, this event is unlikely to occur in normal operating conditions of the device \citep{Hetal10} and won't be investigated here. Further examination of figure \ref{fig:cf} also reveals that the wider the flap the larger the capture factor, for a given width of the channel. This fundamental result, already noticed experimentally by \citet{AL08} and \citet{Hetal10}, reveals a selective behaviour of the sloshing modes with respect to the width of the flap, for which the most powerful resonance occurs with the largest flaps. Now, to what extent can we increase the flap width to exploit this mechanism, in a channel of fixed width? Of course, growth in the structural loads on the device is a heavy limit on the enlargement of the flap \citep{AL08}. Hence we consider the alternative way of decreasing the width of the channel, while holding $w'$ fixed. By decreasing $b'$ so that $w=w'/b'\rightarrow 1$, the width of the channel becomes comparable to the width of the flap. Hence the 3D effects become less noticeable, the geometry approaching that of a 2D device. Figure \ref{fig:2dlim} shows a preliminary analysis of the torque $F'$ with the incident wave period $T'$ for $w=0.99$. 
Here 3D effects are largely inhibited by the body, so that the device shows a ``quasi-2D'' behaviour \citep{YS89}. As a consequence, when $w=O(1)$ the selective behaviour of the sloshing modes is expected to have a weaker impact on the performance of the OWSC, whose maximum efficiency will decrease closer to the theoretical 2D values. This suggests that an optimal layout maximising the resonant actions on the device must lie in between the two extremes $w\rightarrow 0$ (open ocean) and $w\rightarrow 1$ (2D channel). A supplementary investigation of the response of the system to variations of the channel width is then needed. This analysis is currently ongoing and will be disclosed in the near future. 

Finally, recall that resonance of sloshing modes is a linear mechanism. Nevertheless, resonant mechanisms can also excite the second-order transverse modes \citep[cross waves, see][]{SKM87}. A preliminary numerical analysis with a weakly nonlinear Boussinesq model showed nonlinear amplification of the dynamic pressure on the flap to occur near the cross-wave resonant frequencies \citep[see][]{S12}. A nonlinear mathematical model based on the governing equations of Appendix \ref{sec:appA} and on the linear results obtained here is to be developed to investigate whether the nonlinear interactions can further increase the performance of the device.\\

\section{Conclusion}\label{sec:concl}
A mathematical 3D model of an oscillating wave energy converter (i.e. the OWSC) in a channel is developed to investigate the influence of the channel sloshing modes on the performance of the system, in a layout common to many experimental studies \citep[see][]{Fetal05, Wetal07, Hetal10}. The linearised theory, based on the solution of a hypersingular problem in the fluid domain, is satisfactorily validated by comparison with numerical and experimental models \citep[see respectively][]{JO09, AL08}. When the flap is set to oscillations by incident waves, the incoming 2D wave field is modified by the diffraction occurring at the edges of the flap and by the waves radiating from the device. Then a complex 3D wave field develops in the channel, increasing the average efficiency of the OWSC with respect to that of a 2D pitching flap. In addition, this 3D dynamics leads to the excitation of transverse sloshing waves which resonate each at its own cut-off frequency. Transverse modes trapped near the flap are responsible for the spiky behaviour of the hydrodynamic characteristics of the system and further increase the efficiency of the device near resonance.  Finally, parametric analysis reveals a selective behaviour of the sloshing modes, whose resonant capacity depends on the width of the device. In response to the three issues we anticipated in the introductory section, we conclude that (i) resonance of the channel sloshing modes is a linear mechanism and occurs for every width of the flap; (ii) the sloshing modes show a selective behaviour with respect to the width of the flap, that concurs in increasing the performance of the device near resonance; (iii) care should be taken when employing the results obtained in an experimental wave tank to predict the behaviour of the device in the open ocean, especially near resonant frequencies.\\

This work was funded by Science Foundation Ireland (SFI) under the research project ``High-end computational modelling for wave energy systems''. Numerical simulations of \cite{JO09} and experimental data provided by \cite{AL08} in agreement with Aquamarine Power Ltd have been very useful for the validation of the model. Fruitful discussions with Prof. P. Sammarco, Dr K. Doherty and Dr S. Bourdier are kindly acknowledged. The Authors wish to thank the wave research group at Queen's University Belfast for having let them partake in the June 2011 experimental campaign.

\appendix
\section{Nonlinear governing equations}\label{sec:appA}
Consider the geometry of figure \ref{fig:geom} representing the OWSC in the channel as described in \S\ref{sec:anmod}. During the motion, the position of each side of the converter (made up by the bottom foundation and the flap) is at $x'=X'^\pm$, where
\begin{equation}
X'^\pm(z',t')=\left\{
    \begin{array}{ll}
    \pm a', & -h'<z'<-h'+c' \\[2pt]
     -\left(z'+h'-c'\right)\tan\theta'\pm a'/\cos\theta', &  -h'+c'<z'<Z'^{\pm}
  \end{array} \right. .
\label{eq:X'}
\end{equation} 
In the latter, the $+$ ($-$) superscript denotes the right (left) side of the OWSC and $Z'^+$ $(Z'^-)$ is the free-surface elevation along the wetted line of the flap on the right (left) lateral surface (see figure \ref{fig:geom}$a$). Clearly, in (\ref{eq:X'}) the upper equivalence holds for all points on the fixed bottom foundation, while the lower one is valid on the lateral surfaces of the moving flap. As in many gravity-wave problems \citep[see for example][]{ME05} the fluid is deemed to be inviscid and incompressible and the flow irrotational. Hence there exists a potential $\Phi'(x',y',z',t')$ for the velocity field $\boldsymbol{v}'=\nabla ' \Phi'$ such that
\begin{equation}
\nabla '^2\Phi'=0, \quad (x',y',z')\in \Omega',
\label{eq:laplace}
\end{equation}
where $\Omega'$ is the fluid domain. To set up a boundary-value problem governing the behaviour of the fluid in $\Omega'$, the Laplace equation (\ref{eq:laplace}) must be supplied with appropriate boundary conditions. On the free-surface, the kinematic-dynamic boundary condition reads
\begin{equation}
\Phi'_{,t't'}+g\Phi'_{,z'}+\left|\nabla'\Phi'\right|^2_{,t'}+\shalf\nabla'\Phi'\bcdot\nabla'\left|\nabla'\Phi'\right|^2=0,\quad z'=\zeta',
\label{eq:bcfreesurf}
\end{equation}
 where $g$ is the acceleration due to gravity. In (\ref{eq:bcfreesurf}) the free-surface elevation $\zeta'$ is obtained via the Bernoulli equation
 \begin{equation}
-\frac{p'}{\rho}=gz'+\Phi'_{,t'}+\shalf\left|\nabla'\Phi'\right|^2,
\label{eq:bern}
\end{equation}
 where $\rho$ is the water density and $p'(x',y',z',t')$ the fluid pressure. By evaluating (\ref{eq:bern}) at $z'=\zeta'$ and considering only the excess pressure over the atmospheric value, the free-surface elevation is
 \begin{equation}
\zeta'=-\frac{1}{g}\,\Phi'_{,t'}-\frac{1}{2g}\left|\nabla ' \Phi'\right|^2, \quad z'=\zeta'.
\label{eq:freesurf}
\end{equation}
Boundary conditions must also be applied at the solid frontiers delimiting the domain. We require absence of normal flux through the channel walls
\begin{equation}
\Phi'_{,y'}=0,\quad y'=\pm b'/2,
\label{eq:bcwall}
\end{equation}
through the moving sides of the OWSC parallel to the $x'$-axis
\begin{equation}
\Phi'_{,y}=0,\quad X'^-(z',t')<x'<X'^+(z',t'),\, y'=\pm w'/2,
\label{eq:bcflap1}
\end{equation}
and through the impermeable bottom of the channel
\begin{equation}
\Phi'_{,z'}=0,\quad z'=-h'.
\label{eq:bcbott}
\end{equation}
Finally, a kinematic boundary condition allowing only tangential motion along the lateral surfaces of the OWSC $x'=X'^\pm$ is to be applied. Following the reasoning of \citet{SA96} for the mobile gates of the Venice storm barriers, we require $ \de /\de t' \left[x'-X'^\pm \right]=[x-X'^\pm]_{,t'}+\nabla'\Phi'\bcdot\nabla'[x'-X'^\pm]=0$, which yields
\begin{eqnarray}
\Phi'_{,x'}=\left\lbrace\frac{\theta'_{,t'}}{\cos ^2\theta'}\left[-\left(z'+h'-c'\right)\pm a'\sin\theta'\right]-\Phi'_{,z'}\tan\theta'\right\rbrace\nonumber \\
\times \,H(z'+h'-c'),\quad x'=X'^\pm,\: -\frac{w'}{2}<y'<\frac{w'}{2},
\label{eq:bcflap2}
\end{eqnarray}
 where (\ref{eq:X'}) has been employed. In (\ref{eq:bcflap2}) usage of the Heaviside step function $H$ assures absence of flux through the bottom foundation.
 
 The equation of motion of the flap
 \begin{equation}
I'\,\theta'_{,t't'}=T'_g(t')+T'_p(t')+\mathcal{T}'(t')
\label{eq:flapmot}
\end{equation}
 expresses the dynamic equilibrium of torque about the hinge. In (\ref{eq:flapmot}) $I'$ is the second moment of inertia of the flap, assumed to be given, while $T'_g$  and $T'_p$ are respectively the torque due to gravity and to the fluid pressure. Finally $\mathcal{T}'$ is the torque exerted on the flap by the generator, which in general is partly inertial, partly elastic and partly damping \citep[see][]{ME05}. Torque is positive when it makes the flap rotate counter-clockwise, according to the convention adopted for $\theta'$ (see figure \ref{fig:geom}$a$). Since the fluid is considered inviscid and the hinge frictionless, there is no torque induced by viscous tangential stresses. The torque due to gravity is
 \begin{equation}
T'_g(t')=S'g\sin\theta'(t'),
\label{eq:gravtorq}
\end{equation}
where $S'=M'd'$ is the first moment of inertia of the plate. In the latter expression $M'$ is the mass of the flap and $d'$ is the distance between the centre of mass and the hinge \citep[see][]{ME94,SA96}. Both $M'$ and $d'$ are assumed to be given. The net torque $T'_p$ exerted by the fluid pressure $p'$ on the flap is obtained by integrating the product between the unit pressure force $p' \de y'\de z'/\cos\theta'$ and the arm $(z'+h'-c'\mp a'\sin\theta')/\cos\theta'$ respectively on the right and left surfaces of the flap:
\begin{eqnarray}
T'_p=\int_{-h'+c'}^{Z'^+}\int_{-w'/2}^{w'/2}p'\left(X'^+,y',z',t'\right)\frac{z'+h'-c'-a'\sin\theta'}{\cos^2\theta'}\,\de y'\de z' \nonumber \\
- \int_{-h'+c'}^{Z'^-}\int_{-w'/2}^{w'/2}p'\left(X'^-,y',z',t'\right)\frac{z'+h'-c'+a'\sin\theta'}{\cos^2\theta'}\,\de y'\de z'.
\label{eq:prestorq}
\end{eqnarray} 
Summarizing, the complete set of nonlinear differential equations governing the coupled motion of the water and the flap includes: the Laplace equation (\ref{eq:laplace}), the kinematic-dynamic boundary condition (\ref{eq:bcfreesurf}) at the free surface (\ref{eq:freesurf}), the no-flux conditions respectively on the channel walls (\ref{eq:bcwall}), on the sides of the OWSC parallel to $x'$ (\ref{eq:bcflap1}) and on the bottom of the channel (\ref{eq:bcbott}), the kinematic condition on the lateral surfaces of the OWSC (\ref{eq:bcflap2}) and finally the equation of motion  of the flap (\ref{eq:flapmot}).
\section{Solution of the plane radiation and scattering problems}\label{sec:appB}
Solution to the plane radiation and scattering problems of \S\ref{sec:radsca} is sought here with an integral-equation technique based on an appropriate decomposition of the channel Green function.
\subsection{The channel Green function}
Define the Green function $G_n(x,y;\xi,\eta)$ singular at $(\xi,\pm\eta)\in\Sigma$ as the outgoing solution of the Helmholtz equation
\begin{equation}
\left(\nabla^2+\kappa_n^2\right) G_n=0,\quad (x,y)\in\Sigma\setminus(\xi,\pm\eta),
\label{eq:helmG}
\end{equation}
where $\Sigma$ is the 2D fluid domain of figure \ref{fig:geom}($b$), with boundary conditions
\begin{eqnarray}
G_{n,y}=0,\quad &y=\pm 1/2,\label{eq:bcG}\\
G_n\simeq \frac{1}{2\upi}\ln\rho^\pm,\quad &\rho^\pm\rightarrow 0, \label{eq:singG}
\end{eqnarray}
where $\rho^\pm=\sqrt{(x-\xi)^2+(y\mp\eta)^2}$ and $\kappa_n$ are the eigenvalues (\ref{eq:kappan}) of the dispersion relationship (\ref{eq:disprel}), $n=0,1,2,\dots$ The solution of the boundary-value problem (\ref{eq:helmG})--(\ref{eq:singG}) is given in its classical form by \citet{L98}. Here we shall use the decomposition
\begin{equation}
G_n(x,y;\xi,\eta)=-\frac{\img}{4}\left[G_n^{(0)}(x,y;\xi,\eta)+G_n^{(1)}(x,y;\xi,\eta)\right],
\label{eq:Gseries}
\end{equation}
where
\begin{equation}
G_n^{(0)}(x,y;\xi,\eta)=\Ha\left(\kappa_n \sqrt{(x-\xi)^2+(y-\eta)^2}\right) +\Ha\left(\kappa_n \sqrt{(x-\xi)^2+(y+\eta)^2 }\right)
\label{eq:Gn0}
\end{equation}
and
\begin{eqnarray}
G_n^{(1)}(x,y;\xi,\eta)=\sum_{\substack{m=-\infty\\ m\neq 0}}^{+\infty}\left[\Ha\left(\kappa_n \sqrt{(x-\xi)^2+(y-\eta-m)^2}\right)\right.\nonumber\\
\left.+\Ha\left(\kappa_n \sqrt{(x-\xi)^2+(y+\eta-m)^2 }\right)\right]
\label{eq:Gn1}
\end{eqnarray}
with $H^{(1)}_n$ the Hankel function of first kind and order $n\in\mathbb{N}$. Expressions (\ref{eq:Gseries})--(\ref{eq:Gn1}) represent a sum of sources equally spaced along the $y$-axis, due to the mirroring effect of the side walls (as anticipated in \S\ref{sec:goveq}). Note that $G_n^{(0)}$ (\ref{eq:Gn0}) is  singular at $(x,y)=(\xi,\pm \eta)\in\Sigma$, while $G_n^{(1)}$ (\ref{eq:Gn1}) has no poles in $\Sigma$. The simple Green function decomposition (\ref{eq:Gseries}) into a singular part and a converging series will enable us to find appealing fast-convergent semi-analytical solutions to the radiation and diffraction problems. 
\subsection{Solution}
Consider the plane potentials $\varphi^{(R,D)}_n$ solving the system (\ref{eq:helmrad})--(\ref{eq:bcflaprad}), outgoing at large $|x|$. Application of the Green theorem \citep[see for example][]{ME97} to $\varphi^{(R,D)}_n$ and  $G_n$ in the domain $\Sigma$ yields after some algebra
\begin{equation}
\varphi^{(R,D)}_n(x,y)=-\frac{\img}{8}\int_{-w/2}^{w/2} \phijr\, \left[G_{n,\xi}^{(0)}+G_{n,\xi}^{(1)}\right]_{\xi=0}\, \de \eta,
\label{eq:intsol}
\end{equation}
where
\begin{equation}
\phijr=\phijr(\eta)=\varphi^{(R,D)}_n(-0,\eta)-\varphi^{(R,D)}_n(+0,\eta)
\label{eq:pjumpr}
\end{equation}
denotes the jumps in radiation and diffraction potentials from the left to the right side of the flap in the $x$ direction, still unknown. In (\ref{eq:intsol}) the subscript $\xi=0$ indicates the point at which the term in brackets is to be calculated. Application of the boundary conditions (\ref{eq:bcflaprad}) on the wet contour of the plate to (\ref{eq:intsol}) gives
\begin{equation}
-\frac{\img}{8}\left.\frac{\p}{\p x}\int_{-w/2}^{w/2}\left\lbrace
\begin{array}{c}
\Delta\varphi_n^R\\[5pt]
\Delta\varphi_n^D
\end{array}
\right\rbrace\left[ G_{n,\xi}^{(0)}+G_{n,\xi}^{(1)}\right]_{\xi=0} \de\eta\right|_{x=\pm 0} = \left\lbrace 
\begin{array}{c}
V f_n\\[5pt]
A_0\,d_n
\end{array}\right\rbrace, \quad -w/2<y<w/2,
\label{eq:integrodiff}
\end{equation}
which are integro-differential equations for $\phijr$. The structure of (\ref{eq:integrodiff}) would suggest to bring the outer derivative under the integral sign. However, due to the singular behaviour of $G_n^{(0)}$ (\ref{eq:Gn0}), this would lead to a divergent integrand near the poles $\eta=\pm y$. Nevertheless, \citet{MR89} demonstrated that in integro-differential equations like (\ref{eq:integrodiff}) the inversion between the outer derivative and the integral is possible provided the latter is interpreted as a Hadamard finite-part integral $\hint$ \citep{LM01}. By virtue of \citet{MR89}'s theorem, (\ref{eq:integrodiff}) can be rewritten as
\begin{eqnarray}
\hintw\left\lbrace
\begin{array}{c}
\Delta\varphi^R_n\\[5pt]
\Delta\varphi^D_n
\end{array}
\right\rbrace \left.\frac{\partial\left. G^{(0)}_{n,\xi}\right|_{\xi=0}}{\partial x}\right|_{x=0}
 \de \eta+ \nonumber \\
 \int_{-w/2}^{w/2}\left\lbrace\begin{array}{c}
 \Delta\varphi_n^R\\ \Delta\varphi_n^D
 \end{array}
 \right\rbrace \left.\frac{\partial\left. G^{(1)}_{n,\xi}\right|_{\xi=0}}{\partial x}\right|_{x=0}\de\eta=8\img\left\lbrace
 \begin{array}{c}
 Vf_n\\[5pt]
 A_0\,d_n
 \end{array}\right\rbrace.
\label{eq:hypersing}
\end{eqnarray}
The latter are hypersingular integral equations for the jumps in potential $\phijr$ across the flap. The singularity lies in the first term of the left-hand side and is due indeed to the divergent component $G_n^{(0)}$ of the Green function $G_n$ (\ref{eq:Gseries}). To reveal the nature of the singularity, substitute the series (\ref{eq:Gn0}) and (\ref{eq:Gn1}) respectively for $G_n^{(0)}$ and $G_n^{(1)}$ into (\ref{eq:hypersing}). Then perform the double differentiation with respect to $\xi$ and $x$ and use the property $\int_{-\alpha}^{\alpha} f(x)\,\de x = \int_{-\alpha}^{\alpha} f(-x)\,\de x$ \citep[see][\S 3.022]{GR07} to get
\begin{eqnarray}
\hintu \left\lbrace
\begin{array}{c}
P_n(u)\\
Q_n(u) 
\end{array}
\right\rbrace\frac{\Haa\left(\shalf \kappa_n w |v_0-u|\right)}{|v_0-u|}\,\de u\nonumber\\
+\sum_{\substack{
   m=-\infty\\
   m\neq 0}}^{+\infty}\int_{-1}^{1}\left\lbrace
\begin{array}{c}
P_n(u)\\
Q_n(u) 
\end{array}
\right\rbrace\frac{\Haa\left(\shalf \kappa_n w |v_m-u|\right)}{|v_m-u|}\,\de u =\frac{4\img}{\kappa_n}\left\lbrace
\begin{array}{c}
Vf_n\\
A_0\,d_n
\end{array}\right\rbrace,
\label{eq:hypersinglong3}
\end{eqnarray}
for the boundary condition on the wet contour of the flap. In the latter,
\begin{equation}
u=2\eta/w,\quad v_m=2(y-m)/w,\quad \eta,y\in(-w/2,w/2),
\label{eq:varchan}
\end{equation}
while
\begin{equation} 
\left\lbrace\begin{array}{c}P_n(u)\\[5pt]Q_n(u)\end{array}\right\rbrace=\left\lbrace\begin{array}{c}\Delta\varphi^R_n(\shalf wu)\\[5pt] \Delta\varphi^D_n(\shalf wu)\end{array}\right\rbrace=\left\lbrace\begin{array}{c}\Delta\varphi^R_n(\eta)\\[5pt]\Delta\varphi^D_n(\eta)\end{array}\right\rbrace
\label{eq:PnQn}
\end{equation} 
denote the jumps in potential in the new variables. Note that $P_n(u)=P_n(-u)$ and $Q_n(u)=Q_n(-u)$ because of symmetry. Again, as a result of the decomposition (\ref{eq:Gseries}), the singular behaviour of (\ref{eq:hypersinglong3}) is restricted to the finite-part integral of the left-hand side. We must focus on this term to circumvent the singularity. First, expand the Hankel function $\Haa$ according to the series representation \citep[see][\S 8.444]{GR07}
\begin{equation}
\Haa(\shalf \kappa_n w|v_0-u|)=\frac{4}{\img \upi}\frac{1}{\kappa_n w |v_0-u|}+R_n\left(\shalf \kappa_n w |v_0-u|\right),
\label{eq:hanexp}
\end{equation}
where
 \begin{equation}
R_n(\alpha) = J_1(\alpha)\left[1+\frac{2\img}{\upi}\left(\ln\frac{\alpha}{2}+\gamma\right)\right]-\frac{\img}{\upi}\left[\frac{\alpha}{2}+\sum_{j=2}^{+\infty}\frac{(-1)^{j+1}(\alpha/2)^{2j-1}}{j! (j-1)!}\left(\frac{1}{j}+\sum_{q=1}^{j-1}\frac{2}{q}\right)\right]
\label{eq:Rn}
\end{equation}
is the remainder, $J_1(x)$ is the Bessel function of first kind and first order and $\gamma=0.577\:215\dots$ the Euler constant. Then, by making use of (\ref{eq:hanexp}), rewrite the hypersingular integral equations (\ref{eq:hypersinglong3}) as
\begin{equation}
\hintu \left\lbrace\begin{array}{c}
P_n(u)\\Q_n(u)
\end{array}\right\rbrace(v_0-u)^{-2}\,\de u+\left\lbrace\begin{array}{c}\mathcal{K}_n\left[P_n(u)\right]\\
\mathcal{K}_n\left[Q_n(u)\right]
\end{array}\right\rbrace=-\upi\, w\left\lbrace\begin{array}{c} Vf_n \\ A_0\,d_n\end{array}\right\rbrace.
\label{eq:hypersingshort}
\end{equation}
In the latter expression
\begin{equation}
\mathcal{K}_n[f(u)]=\frac{\img\upi\kappa_n w}{4}\int_{-1}^{1} f(u)\left[ \frac{R_n(\shalf k_n w |v_0-u|)}{|v_0-u|}\right. +\sum_{\substack{m=-\infty \\m\neq 0}}^{+\infty} \left. \frac{\Haa(\shalf \kappa_n w|v_m-u|}{|v_m-u|}\right]\, \de u
\label{eq:Kn}
\end{equation}
is an integral function.  Note that as $u\rightarrow v_0\in(-1,1)$, $R_n(\shalf \kappa_n w|v_0-u|)\simeq |v_0-u|\ln|v_0-u|$ from (\ref{eq:Rn}), hence
$\mathcal{K}_n$ has a convergent kernel. 
The singularity in (\ref{eq:hypersingshort}) is finally isolated in the Hadamard integral, whose kernel has a 2nd-order pole at $v_0$. This fundamental result now allows us to solve the hypersingular integral equation (\ref{eq:hypersingshort}) in terms of the jumps in radiation and diffraction potentials, $P_n$ and $Q_n$ respectively. Indeed the Hadamard integral in (\ref{eq:hypersingshort}) admits eigenfunctions that are proportional to the second-kind Chebyshev polynomials $U_p$, $p\in\mathbb{N}$ \citep{LM01}. The latter constitute a complete set of eigenfunctions in the domain $u\in[-1,1]$. Then the analytical form of (\ref{eq:hypersingshort}) itself suggests to seek for solutions of the type
\begin{equation}
\left\lbrace \begin{array}{c}
P_n(u)\\ Q_n(u)
\end{array}\right\rbrace=\left\lbrace\begin{array}{c}V \\A_0\end{array}\right\rbrace\left(1-u^2\right)^{1/2}\sum_{p=0}^{+\infty}\left\lbrace \begin{array}{c} \alpha_{pn} \\ \beta_{pn} \end{array} \right\rbrace U_p(u),
\label{eq:PnCheb}
\end{equation}
where the $\alpha_{pn}$ and $\beta_{pn}$ are unknown complex constants. Note that the position (\ref{eq:PnCheb}) also reflects the asymptotic behaviour of the jump in potential near the tips of the flap $u=\pm 1$, where $P_n, Q_n\simeq \sqrt{(1-u^2)}\rightarrow 0$ and the velocity has a square-root singularity \citep[see][]{N71,LM01}. Now the Chebyshev polynomials $U_p$ satisfy the integral relationship \citep[see][]{PM92}
\begin{equation}
\hintu \frac{\left(1-u^2\right)^{1/2} U_p(u)}{\left(v_0-u\right)^2}\,\de u=-\upi(p+1)U_p(v_0),\quad v_0\in(-1,1),
\label{eq:Chebprop}
\end{equation}
where $v_0=2y/w$ according to (\ref{eq:varchan}) and $y\in(-w/2,w/2)$. Substitution of the series expansions (\ref{eq:PnCheb}) in the relevant hypersingular integral equations (\ref{eq:hypersingshort}) and application of the property (\ref{eq:Chebprop}) yield finally
\begin{equation}
\sum_{p=0}^{\infty}\left\lbrace\begin{array}{c}\alpha_{pn}\\\beta_{pn}\end{array}\right\rbrace C_{pn}(v_0)= -\upi\, w \left\lbrace\begin{array}{c} f_n\\d_n\end{array}\right\rbrace,\quad v_0\in(-1,1),
\label{eq:lineqrad}
\end{equation}
with
\begin{eqnarray}
C_{pn}(v_0)&=& -\upi\,(p+1)\,U_p(v_0)+\frac{\img \upi \kappa_n w}{4}\int_{-1}^{1} \left(1-u^2\right)^{1/2} U_p(u)\nonumber\\
&\times&\left[\frac{R_n(\shalf \kappa_n w|v_0-u|)}{|v_0-u|}\right.+\sum_{\substack{m=-\infty\\ m\neq 0}}^{+\infty} \left. \frac{\Haa(\shalf \kappa_n w|v_0-2m/w-u|)}{|v_0-2m/w-u|}\right]\,\de u.
\label{eq:Cpn}
\end{eqnarray}
The latter can be evaluated by solving numerically the convergent integral in (\ref{eq:Cpn}). For given $n$, expression (\ref{eq:lineqrad}) defines two different linear equations, both valid for any $v_0\in(-1,1)$. The upper equation is for the coefficients $\alpha_{pn}=\alpha_{pn}(\omega)$ of the jump in the plane radiation potential $\varphi^R_n$ across the plate, while the lower equation is for the coefficients $\beta_{pn}=\beta_{pn}(\omega)$ of the jump in the plane diffraction potential $\varphi^D_n$. Theoretically there exists an infinity of points $v_0\in(-1,1)$ where (\ref{eq:lineqrad}) can be evaluated, thus yielding a system of an infinite number of equations  (one for each $v_0$) for an infinite number of unknowns. For (\ref{eq:lineqrad}) to be solved numerically, the unknowns must be truncated to a finite integer number $P<\infty$ and a finite number of evaluation points $v_0=v_{0j}$, $j=0,1,\dots, P$ must be also chosen in the domain $(-1,1)$. \citet{PM92} showed that the fastest numerical convergence for such a system is achieved when the $v_{0j}$ are the zeros of the Chebyshev polynomial of first kind \citep[see][]{GR07}, i.e.
\begin{equation}
v_{0j}=\cos\frac{(2j+1)\upi}{2P+2},\quad j=0,1,\dots,P.
\label{eq:voj}
\end{equation}
Hence (\ref{eq:lineqrad}) reduces to two $(P+1)\times (P+1)$ truncated algebraic systems
\begin{equation}
\sum_{p=0}^{P}\left\lbrace\begin{array}{c}\alpha_{pn}\\\beta_{pn}\end{array}\right\rbrace\, C_{pn}(v_{0j})= -\upi w \left\lbrace \begin{array}{c} f_n\\d_n\end{array}\right\rbrace,\quad j=0,1,\dots,P,
\label{eq:linsysrad}
\end{equation}
for each $n=0,1,2,\dots$ Once (\ref{eq:linsysrad}) has been solved in terms of the $\alpha_{pn}$ and $\beta_{pn}$ for a given modal order $n$, the jump in the $n$-th modal potentials across the plate $\phijr(\eta)$ can be determined with (\ref{eq:PnCheb}) together with the change of variable dictated by (\ref{eq:varchan}) and (\ref{eq:PnQn}). Now substituting $\phijr(\eta)$ into Green's theorem (\ref{eq:intsol}), differentiating the Green function (\ref{eq:Gseries}) inside the integral and summing up all the eigenmodes yield the expressions of the plane potentials. The $n$-th modal plane radiation potential is given by
\begin{eqnarray}
\varphi^R_n(x,y)=-\frac{\img wV}{8} \kappa_n x\,\sum_{p=0}^{M}\alpha_{(2p)n}\sum_{m=-\infty}^{+\infty}\int_{-1}^1 \left(1-u^2\right)^{1/2} U_{2p}(u)\nonumber\\
\times\frac{\Haa\left({\kappa_n\sqrt{x^2+(y-\shalf wu-m)^2}}\right)}{\sqrt{x^2+(y-\shalf wu-m)^2}}\,\de u,
\label{eq:radpotapp}
\end{eqnarray}
where $M=\left\lfloor P/2 \right\rfloor$. In (\ref{eq:radpotapp}) only the even terms survive because the odd terms give no contribution when integrated and summed up over symmetric domains. Finally, the plane diffraction potential is given by
\begin{eqnarray}
\varphi^D_0(x,y)=-\frac{\img w A_0}{8}\,kx \sum_{p=0}^{M}\beta_{(2p)0}\sum_{m=-\infty}^{+\infty}\int_{-1}^{1} \left(1-u^2\right)^{1/2} U_{2p}(u)\nonumber\\
\times\frac{\Haa\left(k\sqrt{x^2+(y-\shalf wu-m)^2} \right)}{\sqrt{x^2+(y-\shalf wu-m)^2}}\,\de u.
\label{eq:difpotapp}
\end{eqnarray}
Here only the $0$-th modal component $\varphi^D_0$ is non-zero; in other words, the diffraction problem only admits the fundamental mode $n=0$. This happens since $d_n=0$ for $n> 0$ (see \ref{eq:Dn}). Hence the $\beta_{pn}$, which are homogeneous solutions of (\ref{eq:linsysrad}) for $n> 0$, must be zero to ensure uniqueness of the non-homogeneous solutions $\alpha_{pn}$ for $n>0$. This is a solvability condition of the coupled radiation-diffraction problems, which substituted into (\ref{eq:PnCheb}) and then into (\ref{eq:intsol})  yields ultimately $\varphi^D_n=0$, $n>0$.
\section{Further relations between radiation and scattering}\label{sec:appC}
In this section we provide some relations between the radiation and the scattering problems discussed in \S\ref{sec:anmod}. First consider the excitation torque $F$ given by (\ref{eq:Td}). The latter depends on $\beta_{00}$, solution of the linear system (\ref{eq:linsysrad}) for the scattering problem. Here we show that that $F$  can also be written in terms of the solutions $\alpha_{00}$ to the linear system (\ref{eq:linsysrad}) for the radiation problem. Consider the two systems of equations (\ref{eq:linsysrad}) for $\alpha_{p0}$ and $\beta_{p0}$ respectively. Since the coefficients $C_{p0}$ are the same in either system, then one must have $\beta_{p0}=\alpha_{p0} d_0/f_0$ to ensure uniqueness of the solution. Hence substitution of the latter expression into (\ref{eq:Td}) and the use of (\ref{eq:Dn}) for $d_0$ yield after some algebra
\begin{equation}
F=-\frac{\img\upi w}{8}A_0\frac{\sqrt{k}\left(2kh+\sinh 2kh\right)^{1/2}}{\cosh kh}\,\alpha_{00}
\label{eq:Tdrad}
\end{equation}
for the complex excitation torque. In (\ref{eq:Tdrad}) $\alpha_{00}$ is given by the solution of the system (\ref{eq:linsysrad}) for the \textit{radiation} problem, which then suffices to describe the dynamic actions on the flap. In addition, a direct relationship between the diffraction torque $F$ and the radiation damping $\nu$ (\ref{eq:raddamp}) can also be found. First, numerical evaluation of the coefficients $\alpha_{0n}$ reveals that $\Imag\left\lbrace\alpha_{0n}\right\rbrace=0$ for all $n>0$. As a consequence, expression (\ref{eq:raddamp}) for the radiation damping simplifies as
\begin{equation}
\nu = \frac{\omega\upi w}{4}f_0\, \Imag\left\lbrace \alpha_{00}\right\rbrace.
\label{eq:nusimp}
\end{equation}
Then by taking the real part of $F$ (\ref{eq:Tdrad}) and making use of (\ref{eq:nusimp}) to express $\Imag\left\lbrace \alpha_{00}\right\rbrace$ in terms of $\nu$, the sought relationship between the radiation damping and the excitation torque is found
\begin{equation}
\nu=\frac{1}{k\,C_g}\left[h-c+\frac{\cosh kc - \cosh kh}{k\, \sinh kh} \right]\,\tanh kh\, \Real\left\lbrace\frac{F}{A_0}\right\rbrace,
\label{eq:nuTD}
\end{equation}
$C_g$ being the non-dimensional group velocity corresponding to (\ref{eq:Cgp}). Expression (\ref{eq:nuTD}) is the counterpart of the 2D Haskind-Hanaoka relation that associates the excitation torque acting on the fixed flap in incoming waves with the radiation potential of the pitching motion of the body \citep[see for example][]{ME05}. According to (\ref{eq:nuTD}) a close link exists between the radiation and the scattering problems, for which they both exhibit the same resonant behaviour, as pointed out in \S\ref{sec:resonance}. Finally, (\ref{eq:nuTD}) has also been employed as a benchmark to double-check the solution of the whole system.
\bibliographystyle{jfm}

\bibliography{biblio}

\end{document}